\documentclass[aip,rsi,preprint,graphicx]{revtex4-1} 
\usepackage[dvipdfm]{graphicx}

\draft 

\begin{document}


\title{Self-resonant Coil for Contactless Electrical Conductivity Measurement under Pulsed Ultra-high Magnetic Fields} 



\author{Daisuke Nakamura$^1$, Moaz M. Altarawneh$^2$, and Shojiro Takeyama$^1$}
\email[]{takeyama@issp.u-tokyo.ac.jp}
\affiliation{$^1$Institute for Solid State Physics, University of Tokyo, 5-1-5, Kashiwanoha, Kashiwa, Chiba 277-8581, Japan\\
$^2$Department of Physics, Mu'tah University, Mu'tah, Karak, 61710, Jordan
}


\date{\today}

\begin{abstract}
In this study, we develop experimental apparatus for contactless electrical conductivity measurements under pulsed high magnetic fields over 100 T using a self-resonant-type high-frequency circuit. 
The resonant power spectra were numerically analyzed, and the conducted simulations showed that the apparatus is optimal for electrical conductivity measurements of materials with high electrical conductivity. 
The newly developed instruments were applied to a high-temperature cuprate superconductor La$_{2-x}$Sr$_x$CuO$_4$ to show conductivity changes in magnetic fields up to 102 T with a good signal-to-noise ratio. 
The upper critical field was determined with high accuracy.
\end{abstract}


\maketitle 

\section{Introduction}
Contactless methods for electrical conductivity measurements have been widely utilized for non-bulk solids such as powder and liquid, or for materials for which it is difficult to obtain an ohmic contact. 
Contactless techniques are also useful in noisy environments during measurements such as the case under pulsed strong magnetic fields\cite{Sakakibara1989RSI}. 
When applying a pulsed magnetic field, discharging electromagnetic noise is one of the barriers for electrical measurements. 
Therefore, a radio frequency (RF) wave with a frequency that is considerably higher than that of electromagnetic (EM) noise was used as a probe signal to filter the noise\cite{Sakakibara1989RSI}. 
In addition, the fast sweep rate of the pulsed magnetic field induces a high inductive voltage in lead wire loops near the contact. 
In ref. 1, a transmission-type configuration was used; the sample was set between a pair of miniature coils, and the RF wave that radiated from one coil was transmitted to the second coil through the sample. 

The use of contactless methods has become more essential in ultra-high magnetic fields above 100 T, which are only generated by destructing magnet-coils, using methods such as the flux compression and the single-turn coil (STC). 
In both cases, a mega-ampere electrical current is instantaneously discharged and it induces a large amount of EM noises. 
The experimental setup in Ref. 1 has been applied to the STC system in magnetic fields of up to 100 T \cite{Sakakibara1989PB}, and subsequently to the electromagnetic flux compression system up to 280 T for electrical conductivity measurements that were aimed at determining the upper critical field of YBa$_2$Cu$_3$O$_{7-x}$ (YBCO) \cite{Sekitani2004PB,Sekitani2007NJP}. 
However, the transmission-type configuration employed above is not always adequate for materials possessing high electrical conductivity, because RF wave transmission is limited by skin depth. 

Kudasov $et al.$ developed a simpler configuration for contactless measurement, which uses only a single coil at an end of the transmission line of the RF wave \cite{Kudasov1998JETP}. 
The reflectivity of the RF wave at the coil depends greatly on the sample that is electromagnetically coupled to the coil. 
This technique is more suitable for samples with high electrical conductivity because of the reflection-type configuration, and the electrical conductivity of FeSi was attempted to measure up to 400 T generated by the chemical explosive flux compression method \cite{Kudasov1998JETP}. 
However, data resolution was poor owing to the limited dynamic range of their instrument. 
Furthermore, the frequency of the RF wave that they used for the measurement was approximately 50 MHz, which is still close to the spectral range of the EM noise generated by the destructing pulsed magnets. 
Therefore, for more precise electrical conductivity measurements above 100 T, we require a measurement system with a higher dynamic range and a higher RF wave frequency.

Recently, in applications to obtain measurements in magnetic fields close to 100 T using non-destructive-type magnets, a resonant-type circuit has been employed for the high-sensitive contactless electrical conductivity measurements. 
Resonant-type circuits with the tunnel diode oscillator (TDO) or the proximity detector oscillator (PDO) were widely used as tools for high sensitivity measurements \cite{Coffey2000RSI,Ohmichi2004RSI,Moaz2009RSI}. 
However, the optimum parameter for the stable operation of TDO and PDO is generally limited to a narrow range. 
In addition, TDO and PDO are significantly affected by the high sweep rate of pulsed magnetic fields \cite{Moaz2012RSI}. 
Therefore, these oscillators are not always suitable for using under destructive pulsed magnetic field. 
On the other hand, a band-stop resonant circuit was recently utilized by one of the authors with a non-destructive-type pulse magnet \cite{Moaz2012RSI}. 
Although a high dynamic range can be obtained using resonant-type circuits, it has not been systematically evaluated how much the resonant spectrum evolves as the sample electrical conductivity changes. 
This is an important issue that needs to be addressed for the quantitative analysis of the measurement system.

Here, we show our proposed contactless measurement system using a band-stop resonant circuit, which achieves stable operation in ultra-high magnetic fields generated by STC destructive-type magnets. 
The frequency response of a probe coil was numerically evaluated. 
The instrument was then applied to measurements of the upper critical field of the cuprate high-temperature superconductor La$_{2-x}$Sr$_x$CuO$_4$ ($x$ = 0.16).

\section{Instrumentation}

\begin{figure}
\includegraphics[width=8cm]{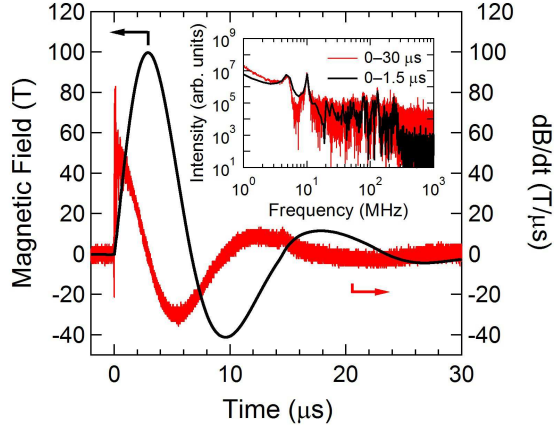}
\caption{\label{Fig1}Waveform showing the magnetic field generated by the single-turn coil system (black) and the signal of the pickup coil, $dB/dt$ (red).
(Inset) Power spectrum of electromagnetic noise detected by the pickup coil at a time between 0-1.5 $\mu$s (black), and between 0-30 $\mu$s (red).}%
\end{figure}

Figure 1 shows a typical magnetic field curve generated by the STC system and the corresponding raw signal from a pickup coil, $dB/dt$. 
Immediately after discharging into a magnet coil, a large electrical noise with a frequency spectrum extending to approximately 300 MHz appeared within 1.5 $\mu$s. 
This initial discharging noise is more apparent when comparing the power spectrum of the pickup coil signal in time from 0 $\mu$s to 1.5 $\mu$s (noisy part, the bold curve in the inset of Fig. 1), and that from 0 $\mu$s to 30 $\mu$s (whole part, the thin curve). 
Therefore, the discharging noise can be filtered out by employing a frequency above the 300 MHz RF wave as a probe signal, and by filtering out the low-frequency discharging noise from the output RF signal.

\begin{figure*}
\includegraphics[width=15cm]{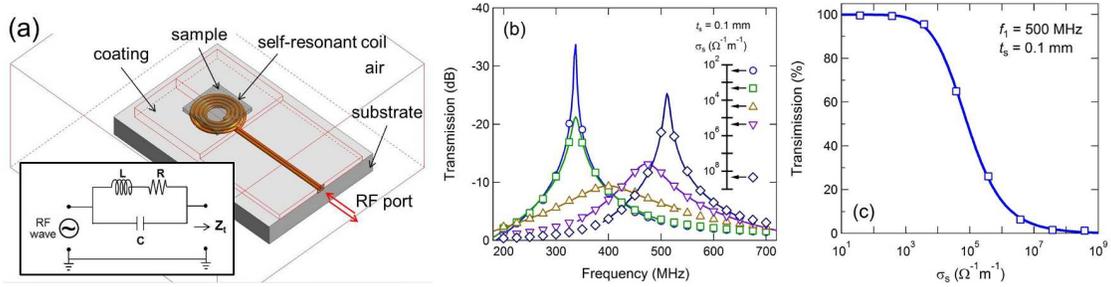}
\caption{\label{Fig2}(a) Model of the SRC used in the electromagnetic analysis. 
The inset is the equivalent circuit of the SRC. 
(b) Calculated power spectrum of a self-resonant coil for different values of $\sigma_s$ (3.8 $\times$ 10$^2$, 3.8 $\times$10$^3$, 3.8 $\times$ 10$^4$, 3.8 $\times$ 10$^5$, 3.8 $\times$ 10$^8$ $\Omega^{-1}$m$^{-1}$). 
(c) Normalized transmission as a function of $\sigma_s$ at $f_1$ = 500 MHz calculated for $t_s$ of 0.1 mm.}%
\end{figure*}

The probe coil was wound 10-12 turns with a flat spiral-type structure, as shown in Fig. 2(a). 
The sample was placed directly on the coil in order to improve EM coupling. 
The outer diameter of the probe coil was set to be approximately 1 mm, which is the size at which the homogeneity of the magnetic field is maintained at approximately 0.3 \% when the probe coil is settled at the center of a STC with a bore of 14-16 mm. 
The probe coil is regarded as a band-stop equivalent circuit (inset of Fig. 2(a), also see ref. 9), whose complex impedance $Z_t$ is given by
\begin{equation}
|Z_t | = \left( \frac{R^2 + \omega^2 L^2}{(1-\omega^2 LC)^2 + \omega^2 R^2 C^2} \right)^{0.5}
\end{equation}
\begin{equation}
arg (Z_t) = \tan ^{-1} \left[ \frac{\omega L (1-\omega^2 LC) - \omega R^2 C}{R} \right]
\end{equation}
with a resonant frequency $2\pi f_{\text{res}} = (1/LC)^{0.5}$. 
Hereafter, we called our method the self-resonant method and the probe coil the self-resonant coil (SRC). 
A substantial change in the power spectrum of a SRC is induced by an electromagnetically coupled sample, especially near $f_{\text{res}}$. 
In other words, the dielectric property of a sample attached to the SRC is determined by finding changes in the resonant spectra. 

The frequency response of the self-resonant method was numerically evaluated. 
The power spectrum of a SRC is calculated by performing an electromagnetic analysis using commercially available finite element analysis software (Femtet, Murata manufacturing Co, Ltd.). 
Figure 2(a) illustrates the SRC and materials around the sample holder modeling the real probe. 
The SRC is modeled by a spiral copper wire and connected to the input and output RF ports via copper wires aligned in parallel. 
The substrate and the coating of the SRC are composed by Stycast1266. 
The sample thickness $t_s$ is set to be 0.1 mm. 
The power spectrum was calculated for several values of the electrical conductivity of the sample, $\sigma_s$ . 
In the calculation, the imaginary part of the electrical conductivity was ignored, similar to the calculation performed in the case of the transmission method \cite{Sakakibara1989RSI}.

The calculated SRC power spectrum was shown in Fig. 2(b). 
The increase in $\sigma_s$ resulted in a high-frequency shift of the resonant peak. 
The peak height becomes minimal at $\sigma_s$ = 3.8$\times$10$^4$ $\Omega^{-1}$m$^{-1}$, where skin depth $\delta$ becomes comparable to $t_s$ ($\delta \sim$ 0.1 mm at $f_{\text{res}}$). 
The spectral changes in the power spectrum cannot be traced in pulsed magnetic fields because of the single-shot nature of the magnetic field with the duration of a few microseconds. 
Hence, the transmitted RF signal with fixed frequency was detected during the pulsed magnetic field duration time. 
For example, if we set the input frequency $f_1$ of the RF wave to be 500 MHz in Fig. 2(b), the decrease of $\sigma_s$ induces a large increase in the transmission amplitude of the RF wave. 
This is shown in Fig. 2(c) as a function of $\sigma_s$ . 
The intensity remains constant up to 10$^3$ $\Omega^{-1}$m$^{-1}$, and then, there is a significant drop between 10$^3$ and 10$^7$ $\Omega^{-1}$m$^{-1}$, where the SRC response to the sample is most efficient. 

\begin{figure}
\includegraphics[width=10cm]{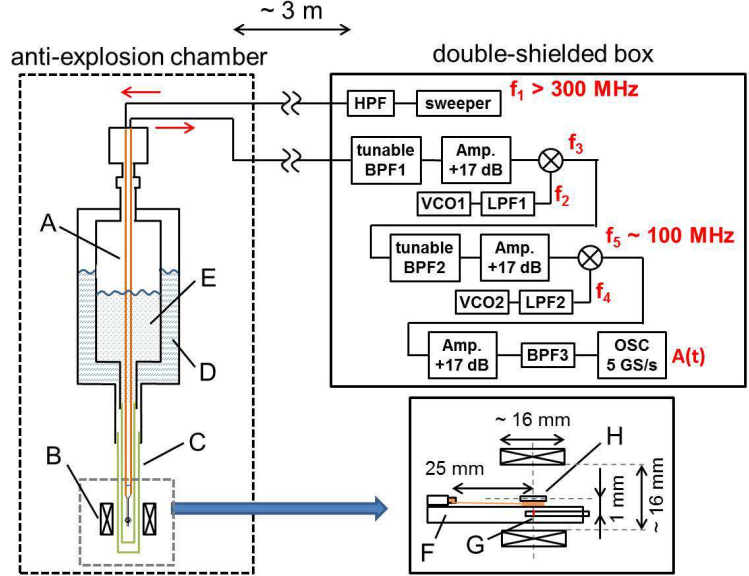}
\caption{\label{Fig3}Schematics of the experimental apparatus. 
A: the semi-rigid coaxial cable in the measurement probe, B: the single-turn coil, C: G-10 tail of the cryostat, D: Liq. N$_2$ dewar ($\sim$ 0.7 liter), E: Liq. He dewar ($\sim$ 0.4 liter), F: bottom part of the measurement probe, G: the pickup coil, H: sample and the self-resonant coil. }%
\end{figure}

The experimental apparatus is illustrated in Fig. 3. 
The vertical STC (V-STC) system is employed for the generation of a pulse magnetic field over 100 T \cite{Miura2003JLTP}. 
A liquid helium bath cryostat \cite{Takeyama2012JPSJ} was settled in the 14 mm-bore of the STC. 
The measurement probe was directly immersed in liquid helium. 
A semi-rigid cable of 1.19 mm-diameter (COAX Co. Ltd., SC-119/50) was used to guide the RF waves in and out from 25 mm away from the center of the STC. 
The SRC is then connected by a pair of 60 $\mu$m-thick parallel copper wires. 
These are shown in the enlarged illustration in Fig. 3. 
The SRC was wound using a polyamide-imide enameled 0.06 mm-diameter copper wire (AIW TOTOKU Electric Co. Ltd.), and was glued to the sample stage by GE7031 varnish. 
The sample was also attached on the SRC by GE7031 varnish. 
In order to prevent the induction of a large voltage by the pulse magnetic field, the plane of the SRC was placed parallel to the direction of the magnetic field. 
A pickup coil wound around the G10 rod for monitoring the magnetic field was inserted adjacent to the sample. 

The input RF signal was generated by a sweeper (Agilent N5181A) placed inside the double-shielding box. 
The input frequency $f_1$ was chosen such that it is close to $f_{\text{res}}$ at each measurement temperature. 
The typical input power of the RF wave is -5 dBm, which is much smaller than that used in the transmission-type measurement. 
The output RF signal returns to the double-shielding box, and is amplified and filtered by the double-stage superheterodyne circuit. 
Because $f_1$ was tuned at different temperatures and adjusted for samples, the tunable band-pass filters (Trilithic, Inc.) were used as BPF1 and BPF2 in Fig. 3. 
After the filtering and amplification, the frequency of the RF signal was down-converted by a mixer. 
We used a homemade voltage-controlled oscillator (VCO) circuit as the source of the intermediate frequency of the mixer. 
For example, when $f_1$ = 750 MHz, the frequencies of VCO1 and VCO2 were set to be $f_2$ = 400 MHz and $f_4$ = 250 MHz, respectively. 
Low pass filters (LPF1 and LPF2 in Fig. 3) after VCOs were used to eliminate the higher harmonic wave. 
Finally, the frequency of the RF wave was converted to approximately 100 MHz ($f_5$ in Fig. 3), and a high-resolution digital oscilloscope (Recroy, WaveRunner 44Xi) was used for the data acquisition. 
Using the band pass filter after the frequency down-conversion, narrow-band filtering generally becomes possible. 
The power of all of the active components in the super-heterodyne circuit was supplied by a 12 V lead acid battery.

In the frequency-converted RF signal ($A(t)$ in Fig. 3), the time-dependent amplitude $a(t)$ and phase $\phi (t)$ were calculated by the following numerical lock-in method \cite{Tukey1961}. 

\begin{eqnarray}
&A(t) = a(t) \cos (2\pi f_5 t + \phi (t)) + b(t) \nonumber \\
 &\hspace{5cm} \to \left\{ 
\begin{array}{ll}
a(t)  = 2 \sqrt{F^2[A(t) \cos 2\pi f_5 t ] + F^2 [A(t) \sin 2\pi f_5 t ] ]}\\
 \phi (t) =- \text{arctan} \left( \frac{F[A(t) \sin 2\pi f_5 t]}{F[A(t) \sin 2\pi f_5 t}  \right) 
\end{array}
\right. 
\end{eqnarray}

where $F$ means the numerical low pass filter. 
After the numerical lock-in process, the time-dependent offset signal $b(t)$ in $A(t)$ is excluded.

\section{Application}
The system described above was applied to measure the conductivity of the cuprate high-temperature superconductor La$_{2-x}$Sr$_x$CuO$_4$ (LSCO) in ultra-high magnetic fields. 
The upper critical magnetic field ($B_{c2}$) of cuprate superconductors is expected at very high magnetic fields around 100 T at very low temperatures. 
In the present study, the optimally-doped ($x$ = 0.16) single crystal of LSCO \cite{Adachi2009JPSJ} was used as a measurement sample, and its $T_c$ was determined to be 36.7 K from the magnetic susceptibility. 
The tabular shaped sample parallel to the ab plane was prepared, and its dimension was approximately 1 $\times$ 1 $\times$ 0.05 mm$^3$. 
The magnetic field was applied along the ab plane ($B//ab$). 

\begin{figure}
\includegraphics[width=9cm]{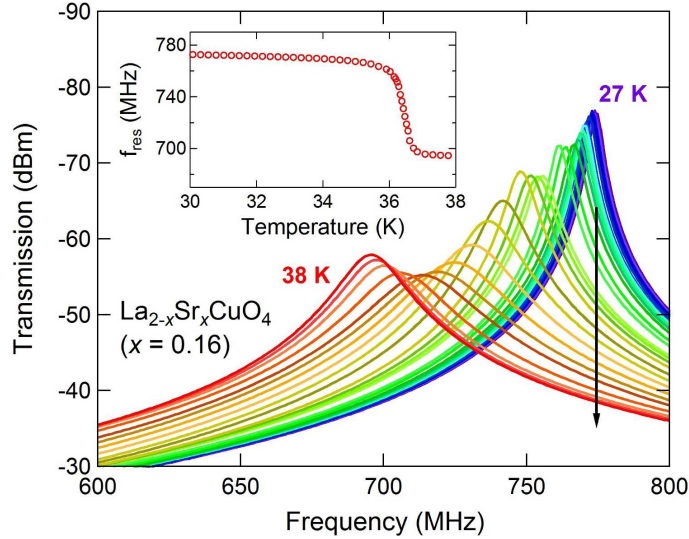}
\caption{\label{Fig4}Temperature dependence of the power spectrum of La$_{1.84}$Sr$_{0.16}$CuO$_4$ around $T_c$. 
Inset shows the temperature dependence of the resonant frequency. }%
\end{figure}

Figure 4 shows the power spectrum of the SRC with a sample LSCO at different temperatures ranging from 38 K down to 27 K. 
At low temperatures below $T_c$ = 36.7 K, the large increase of $\sigma_s$ results in the enhancement of the electromagnetic induction against the SRC, which reduces the effective inductance of the SRC. 
As a result, $f_{\text{res}}$ increases as is plotted in the inset of Fig. 4. 
We observe that there is an abrupt change in the $f_{\text{res}}$ around $T_c$, where the S-N transition is manifested.

\begin{figure}
\includegraphics[width=9cm]{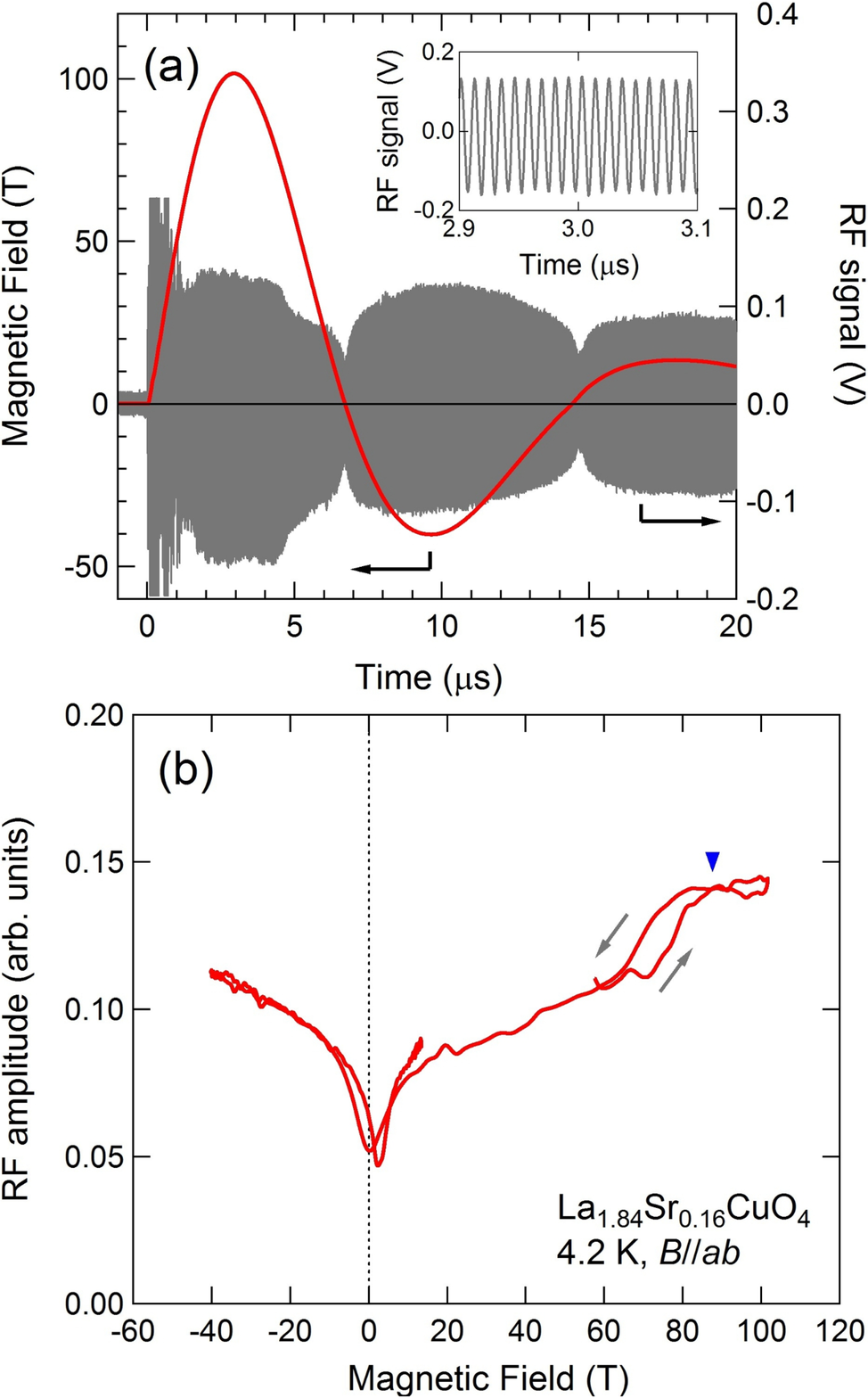}
\caption{\label{Fig5}Electrical transport characteristics of La$_{1.84}$Sr$_{0.16}$CuO$_4$ under an ultra-high magnetic field up to 102 T. 
(a) Frequency-converted RF signal and the magnetic field curve. 
The inset is an enlarged view of the RF signal. 
(b) Amplitude of RF signal as a function of the magnetic field.}%
\end{figure}

Figure 5 shows the result of the measurement subjected to pulsed magnetic fields up to 102 T at 4.2 K. 
$f_1$ was chosen to 780 MHz close to $f_{\text{res}}$ at 4.2 K. 
Therefore, at the S-N transition, the drastic decrease of $\sigma_s$ causes a low-frequency shift of the resonant peak, which in turn induces an abrupt increase in the RF transmission amplitude, as indicated by the downward arrow in Fig. 4. 
Figure 5(a) shows a plot of an output RF signal $A(t)$ from the SRC in a pulse magnetic field duration (a gray line) together with the time profile of the pulsed magnetic field (a solid line). 
The inset of Fig. 5(a) is a plot with the time scale expanded to improve the view of $A(t)$. 
The sinusoidal wave with good S/N ratio is noted. 

In Fig. 5(b), the amplitude $A = |A(t)|$ of the output signal is analyzed by the numerical lock-in method, and is plotted as a function of the magnetic field. 
The amplitude $A$ increases rapidly upon applying the magnetic field, and then at around 10 T, it shows a gradual increase until 65 T. 
Then, there is a hysteresis jump (shown by arrows) until saturation above 85 T. 
The increase of the resistivity in magnetic fields arises from the local suppression of the superconducting order parameter in the penetrating quantum flux core. 
Note that the data in the up-sweep and down-sweep of the pulsed magnetic field almost coincide at the region of low magnetic fields (-40 T$<B<$ 10 T), exhibiting less of the overheating effect arising from the eddy current induced by the fast sweep rate of the magnetic field. 
A starts to show saturation at approximately 85 T (indicated by a down triangle), at which the magnetic field is defined as $B_{c2}$. 
A first-order-type phase transition is implied in the hysteresis observed near $B_{c2}$. 
A transition of the first order is expected in the case when the S-N transition is caused by the Pauli paramagnetic effect. 
A detailed discussion will be presented in another article. 

\section{Conclusion}
Using the SRC, we developed a contactless measurement system for the electrical transport properties of samples under ultra-high magnetic fields over 100 T. 
From the frequency response of the self-resonant method calculated by the finite element analysis, we found that the self-resonant method is suitable for measuring samples with a wide range of electrical conductivity (especially for high conductive materials). 
The instrument developed here was demonstrated to show the measurement of the upper critical field of a cuprate high-$T_c$ superconductor LSCO with a sufficiently good S/N ratio in a magnetic field of up to 102 T.

%

\begin{acknowledgments}
We would like to thank Prof. Y. Koike and Prof. T. Adachi for providing the LSCO sample. 
We also appreciate the contribution of Prof. Y. H. Matsuda, with whom we had fruitful discussions.
\end{acknowledgments}



\end{document}